# Aiding the Visually Impaired: Developing an efficient Braille Printer

Anubhav Apurva
Department of Computer and Communication Technology
Manipal Institute of Technology
Manipal University
India
anubhav.apurva@learner.manipal.edu

Anupam Misra
Department of Electrical and Electronics Engineering
Manipal Institute of Technology
Manipal University
India
anupammisra1995@gmail.com

Palash Thakur
Department of Electrical and Electronics Engineering
Manipal Institute of Technology
Manipal University
India
palashthakur1@gmail.com

*Abstract*— With the large number of partially or completely visually impaired persons in society, their integration as productive, educated and capable members of society is hampered heavily by a pervasively high level of braille illiteracy. This problem is further compounded by the fact that braille printers are prohibitively expensive – generally starting from two thousand US dollars, beyond the reach of the common man. Over the period of a year, the authors have tried to develop a Braille printer which attempts to overcome the problems inherent in commercial printers. The purpose of this paper, therefore, is to introduce two prototypes – the first with an emphasis of cost-effectiveness, and the second prototype, which is more experimental and aims to eliminate several demerits of Braille printing. The first prototype has been constructed at a cost significantly less than the existing commercial Braille printers. Both the prototypes of the device have been constructed, which will be shown.

*Keywords*— printer, braille, servos, extrusion, thermoplastic, embossment

## I. Introduction

The number of visually impaired people in India has been increasing steadily over the past twenty years. Estimated to be 24.1 million by 2010, and 31.6 million by 2020, the vast numbers of blind people in India is a great source of concern. Out of a 2004 survey taking 72,044 visually impaired individuals, 71% were found to be illiterate, and 84.6% reside in rural areas [1]. A factor that further makes this problematic is the fact that the cost of Braille printers lies from $2000 to $5000, with large volume Braille printers costing anywhere from $10000 to $80000 [2]. Hence, the need for a device to aid this marginalized section of society is pressing and inarguable. Therefore, the primary focus of this work is to introduce two Braille printing technologies, the working of whom will be explained in the following sections. Section 2 focuses on related patents in the field. Section 3 is the literature review section, with a focus on current trends in the field. The fourth and fifth sections of this paper deal with the first prototype and its basic principle. Section 6-8 deal with the problems present in the first prototype and the development of a second prototype.

A 1996 study [3] analyzing employability and education levels of congenitally visually impaired people with respect to braille versus print as their original reading medium, concluded that those who learned to read using Braille were more employable than those who were not. Through reducing problems in modern braille printing, the team hopes that these improvements may be used to develop Braille devices which can alleviate poverty due to widespread illiteracy amongst the visually impaired populace.

## II. Related Patents

Current braille printers rely on printing techniques mostly reliant on embossment technologies. Multiple designs for printing solenoids have been patented. One of the most commonly cited is a technology patented in 1980[4] - the printing technology is comprised of a base member having a plurality of printing pins arranged in a row. A semicircular printing plate is pressed against the paper through a driving mechanism, and pins corresponding to the character to be printer are pressed downward through the excitation of solenoids.

Another patent deals with a solenoid design for superior Braille printing[5]. A more recent patent deals with an embossment based portable Braille writing system, with a system of keys and corresponding embossment pins being driven into the paper by the corresponding key-press.[6]

Another relevant technology is a 1971 patent (US3598042A) [7], which is also embossment based. Although most Braille printing related literature are embossment based, there have been proposals to print using lithography, as seen in a 1978 patent proposal [8].

## III. LITERATURE REVIEW AND TRENDS IN ASSISTIVE TECHNOLOGIES

This section will deal with a brief overview of recent trends and developments in reading assistive technologies for the blind, particularly pertaining to Braille reading.

Lita, Mazare et al. [9] have built a module (based on a PIC 16F877 microcontroller) to emboss Braille print – the module is meant to be an educational tool for the teaching of Braille. The module appears to be quite fast and efficient – however the costliness of the module is highlighted by the fact that it uses solenoids for printing and 6 stepper motors for positioning of the print head.

Another, less recent, Braille printing implementation by Kocidek, Wiqcek et al. [10] has an interesting proposal – the printing medium used is not paper, but thermoplastic sheets. The printing system is thermal based, and dots are imprinted or erased by local application of heat, which causes embossments or depressions respectively. However, the obvious disadvantage here is the availability of such print media, as well as the fact that the plastic film used has a low heat deflection temperature (40°C). Therefore, the technology is not viable for developing markets or areas where the temperature regularly exceeds the threshold.

While Braille has been the standard for reading material for over a century amongst the blind populace, Braille suffers from a serious limitation – it is unable to convey or represent pictorial information. Gupta, Balakrishnan, and Rao have highlighted the need for tactile graphics to convey this information to the visually impaired [11]. Tactile modality invariably results in the piece-meal acquisition of information by the reader. For Braille, where the reader reads text along a predefined direction, this is an acceptable condition. However, for graphics, this creates complications with preserving the semantic meaning of diagrams, as tactile touch has a limited field of view.

Nadeem, Shaikh et al. carried out a comparative study of Braille refreshable displays [12] – putting forward three main categories of Braille refreshable displays – Linear, Rotating, and Panda displays. Linear and to some extent, Rotating displays suffer from a major disadvantage, that of being prohibitively expensive due to the cost of mechanical actuating devices. Such devices also suffer from a high rate of wear-and-tear, making them unaffordable for developing nations. However, in our view, the Panda display, based on electro-cutaneous stimulations to give the impression of Braille dots, is deserving of further investigation. The Panda display boasts of no moving parts, and a successful implementation could potentially be highly cost-effective. Unfortunately, the nature of the display (sending information via delivering small electric shocks) means that a comprehensive round of testing is required to meet stringent safety standards – the technology does not appear to have been tested in this regard yet.

An interesting implementation of Braille displaying has been carried out by Velasquez, Preza, and Hernandez [13]. This device aims to overcome the disadvantages inherent in conventional relay and lever mechanisms (as used in the Linear and Rotating displays mentioned above). The device uses a multi-layered Braille cell primarily composed of piezoelectric linear miniature motors as the actuating medium.

An extension and application of extrusion technology (as used in the second printer prototype) can be seen in a University of Colorado, Boulder [14] research project where the authors have used tactile graphics (using a 3D printed research book) to aid visually impaired young children.

## IV. BASIC PRINCIPLE OF THE FIRST PROTOTYPE

### A. Aim and methodology of the first prototype

The primary target of the first prototype was to achieve Braille printing in the most inexpensive way possible, that is, the printer's primary aim was to achieve the most economical printing possible, as a proof of concept against expensive commercial printers. Consequently, certain compromises were made.

### B. Components of the Printer

The printer is effectively composed of three modules, each of which is comprised of further sub-modules. These three modules are:

i) The outlying structure

ii) The Printer module

iii) The Controller module

The outlying structure forms the base for the printer, onto which the entire apparatus rests, as well as the chassis of the printer which acts as the base for the print heads.

The Printer module (denoted hereafter by "P- module") is comprised of a development board (Atmel ATMega 328p based) which controls print heads, driven by servomotors. The chief function of the P-module is to receive the input data from the computer to which it is connected, and perform the corresponding printing function onto the input paper.

The Controller module (denoted hereafter by "C-module") is comprised of a second development board, connected to servomotors, coupled with gears. The chief function of the C-module is to receive a signal from the P-module following the printing of a line and to bring about a "Reset" condition, so that the next line can be printed.

### C. Operations of the Printer

There are two ways in which the printer can be operated. Firstly, as a typewriter, i.e., the printer prints text as it is being typed on the computer; and secondly, as a printer, where the entire body of text is given as input and the printer performs the intended operation. When the printer is being utilized as a typewriter, the input text can be transmitted to the P-module (connected to the computer) by entering in text through the program's serial input window. When the printer is being used as such, the input is given through a .txt file sent through the COM port of the PC to the development board utilizing a C++ program, which is then printed.

*D. Generalized working of the Printer*

Analyzing the working of the printer from a top-down general approach, the printer's operation can be broken down as follows:

i) The PC sends the text input to be printed to the P-Module. Each text character is coded to its corresponding Braille character in the program being run by the P- module [15].

ii) The P-module receives the command, prints one character by controlling the print heads.

iii) After printing one character, the P-module then serially transmits a signal to the C-module. This transmission is accomplished by configuring two pins as TX and RX pins.

iv) Upon receiving the signal, the C-module moves the print heads by one character spacing.

v) The above steps are repeated until a fixed number of characters per line are printed.

vi) Once the line has been printed, the P-module transmits another signal to the C-module that signals the C-module to revert the print head back to the initial position, and to push the paper by one line.

vii) Processes (i) to (vi) are repeated until the entire document has been printed.

In the case of a document which requires multiple pages to be printed, the user must insert a fresh page into the printer feed and manually resume the printing process, whereupon the printer will resume printing.

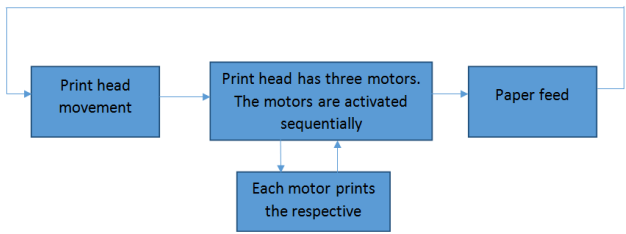

Fig. 1. Block Diagram for the first prototype

## V. FIRST PROTOTYPE

The prototype is composed of the same three basic components and operates on the same general working as stated in the previous section. Hence, detailing on the working of the printer through the example of the prototype is given in the following sections.

*A. Structure*

The outlying structure is composed of plywood, and consists of a platform (30x50 cm) which acts as a base for the printer. There is a metal chassis – two steel plates of height (9 cm) separated by a distance of (24.5 cm) that act as the passageway for the paper. The plates are mounted onto the base on a separate (2 cm) raised base of plywood, and fixed by L shaped axle clamps. Through the chassis is fitted one set of metal axles of diameter (0.3 cm) - near the top at a height of (8.5 cm). Each axle in the pair is separated by (3.8 cm). There is a rectangular structure, made of plywood, fixed onto the base, which houses the printer arrangement.

The function of the axles is to facilitate paper movement, as well as to act as a base for the print heads. There are two cardboard platforms (one on the lower set, the other on the upper set) fixed to the axles for this purpose.

*B. P-module and working of the print heads*

The P-module (or Printing module) is composed of a development board, connected to the PC. To this board are connected three micro servomotors (model SG90) of torque 1.8 kg-cm, capable of 180° rotation. These servomotors form the basis for the print heads. To each servomotor, there are two needles diametrically attached by a distance of (0.6 cm) at a plane perpendicular to the shaft axis of the motor. Two motors are arranged on opposite ends on the lower cardboard platform for printing of the upper and lower pairs of dots for each character. The upper platform has the third servomotor (with a longer pair of needles diametrically attached) arranged such that it is able to print the middle hole pair.

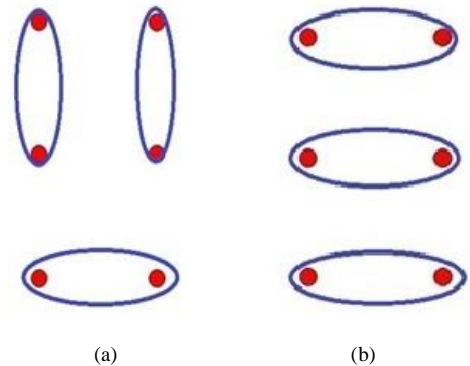

Fig. 2. Allignment of the servos

From Fig. 2.(b) it can be observed that the second (and final design) enables the print head to be more compact overall as it allows all three servomotors to be placed longitudinally. The print head needles are channeled through plastic pipes of diameter (0.4 cm) fitted through perforations in the cardboard platform. Each text character is coded to a specific hole pattern (the corresponding character in Braille). Accordingly, each servomotor rotates clockwise or anticlockwise, to create the right or left hole respectively. Thereby, the Braille character is printed.

*C. The C-Module and operation of the "Reset" condition*

The C-module (or Controller module) is composed of a second development board, to which two servomotors (model FS5106R) of torque 6 kgcm are connected. These servomotors are capable of continuous rotation, and have metal gears. One servomotor is fixed on top of the rectangular plywood structure. The gear of the servomotor is connected to the print head arrangement by way of a rack and pinion arrangement. The function of this servo, therefore, is to move the print heads laterally by one character spacing, in order to print the next character. After a set number of characters have been printed (this is accomplished by incorporating a counter in the code), the servo performs the "Reset" condition and moves the print head arrangement back to the initial position.

The second servomotor, meanwhile, is connected to a gear axle situated at the back of the printer. Its purpose is to rotate by 1 cm line space after each line and thereby complete the "Reset" condition by pushing the paper forward by one line.

*D. Cost Analysis of the Prototype and Performance*

One major advantage of this printer can be seen in its cost-effectiveness. The plywood base of the prototype cost approximately ₹200, while the combined cost of the two development boards being utilized was ₹1200. The half rotation micro servomotors cost ₹300 each while the two full rotation servomotors cost ₹600 each. All in all, the entire cost of the prototype approximates to ₹4000, which is truly minuscule when compared to the cost of a commercial Braille printer ($2000 - $5000).

The prototype has been demonstrated to be able to print one Braille character in 1-3 seconds, and to be able to print a 24 character line in approximately 25-30 seconds, depending on the characters being used.

*E. Power requirements*

Unlike conventional Braille printers, the prototype does not require external AC power to operate. The P-module is operated through the connection with the PC, and the motors are powered either through a 9V DC battery, or by connecting a wall socket.

## VI. Problems in the first prototype and Need for the Second Prototype

As the primary focus of the first prototype was to achieve a form of Braille printing under rigorously frugal conditions, compromises were made. Speed and accuracy of printing, as well as robustness of the structure was not a high priority. Therefore, in spite of its advantage of economical printing, the crude state of the prototype ensured its status as merely a proof of concept. However, in terms of achieving the basic target of cost-effectiveness, the prototype was a success.
Keeping these issues in mind, development on the second prototype was started. Currently, progress is still ongoing on the second prototype to make it a standalone system, although the primary objectives have been achieved as of the writing of this paper.

## VII. Basic Principle of the Second Prototype

In view of the shortcomings of the first prototype, the second prototype was developed. Therefore, the aims of the second prototype were-

i) To achieve a form of Braille print which is durable and long—lasting, as well as quick.
ii) To be structurally robust, as well as mechanically reliable.
iii) To achieve printing on regular inexpensive lower GSM paper, as opposed to commercial printers which require specialized high GSM paper.

While developing the second prototype, the team pivoted to a new technology; that of extrusion. While the previous prototype relied on imprinting dots onto the paper itself, and commercial printers rely on embossing dots onto the paper itself, this printer essentially extrudes dots onto the input paper by using quick setting, inexpensive, thermoplastic sticks as the input. The thermoplastic stick is heated and its temperature is constantly monitored using a control circuit.

*Adhesive Composition*

Domestic hot glue adhesives (like the one used in the second prototype), show a great deal of variety in composition, bond strength, and viscosity. The thermoplastic adhesive being used for the printer is a typical EVA (ethylene-vinyl acetate) copolymer hot-melt adhesive.

## VIII. Second Prototype and Advantages

The role of each block (Fig. 3.) in the overall working of the printer is as follows-

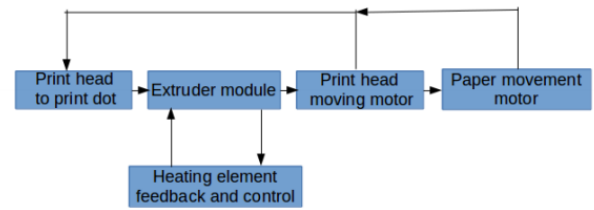

Fig. 3.  Block Diagram for the second prototype

i) Print head – This contains the input thermoplastic stick which will be melted by the extruder module.
ii) Extruder module – The extruder module is joined to the Printer head and moves along with it. Its purpose is to melt the input stick and extrude it on the target paper.
iii) Motor to move the print head – This moves the print head to print the next dot as the dots are being printed.
iv) Heating element feedback and control mechanism – To heat the thermoplastic stick, there is a heating element in place. To prevent the heating element from overheating the thermoplastic, a control mechnism is in place using a temperature sensor which maintains the temperature at 125 degrees Celsius.

v) Paper driving motor – Once a line of dots has been printed, the paper movement motor moves the paper by one line spacing.

Based on these ideas, a CATIA model and 3D rendering was developed, which was later converted into a full prototype.

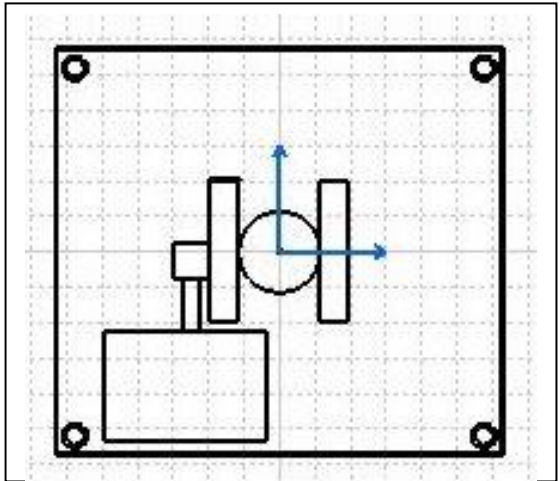

Fig. 4. Top View of Print Head

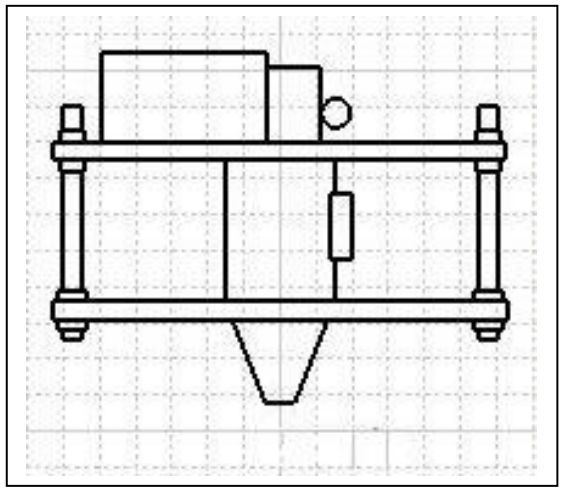

Fig. 5. Front View of Print Head

## IX. COMPARISON OF THE TWO PROTOTYPES

| Criterion | 1st prototype | 2nd prototype |
|---|---|---|
| Printing style | Character based | Line based |
| Cost | ₹4000 | ₹8000 |
| Printing Method | Embossment based | Extrusion based |
| Running cost | None | ₹20 per stick of thermoplastic |

Table 1. Comparison between the prototypes

One important difference between the first prototype and the second in terms of printing is that the former prints one character at a time, while the latter prints a line of dots at a time. It is evident from this that the second prototype would require three lines of dots to print a complete row of characters.

Due to printing thermoplastic dots through extrusion onto the paper, advantages of the new technology include an ability to print on any type of paper, not just high GSM paper, as well as faster and more efficient printing. Obviously the extrusion based 2nd prototype requires thermoplastic sticks as input. Fortunately, these sticks are quite inexpensive, and one can print up to 10 pages per stick.

The new technology, while still in its experimental stages, shows promise, especially in comparison to the older prototype. The Braille print is long lasting, sustainable, and durable print. In comparison, printers commonly utilize embossments which flatten over time, resulting in a printed product which has a relatively short lifespan.

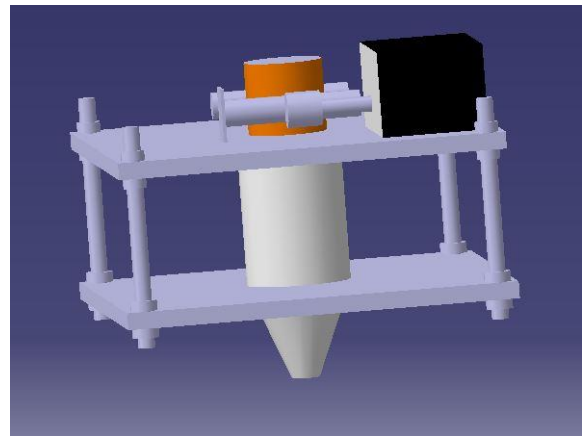

Fig. 6. 3D rendering of the Print Head

## X. Possible Improvements and Future Scope

It is to be noted that the two technologies developed through the two prototypes are not meant to be a complete Braille printing solution. However, work on the second prototype is ongoing as of the writing of this paper where the speed and reliability of the printing mechanism is being worked upon.

As far as the future scope of the printer goes, many applications can be thought of. One major extension that the printer could be coupled with is that of a speech-to-text software; the user could speak into a voice input device, and the software could convert it to text format – ready to be printed by the printer.

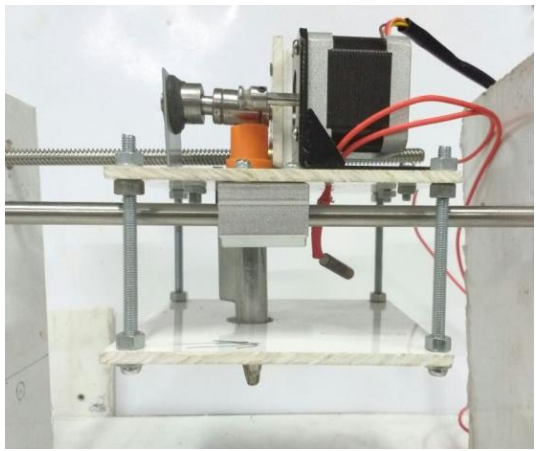

Fig. 7. The print head of the second prototype

## XI. Conclusion

The two printer prototypes were developed with highly specific goals in mind. It is evident that the two prototypes were for primary research purposes and wholly experimental in nature. Therefore, a complete Braille printing solution must keep this in consideration. However, the research was successful insofar as it serves to prove that a viable, inexpensive, and complete solution to Braille printing is possible, and several limitations of commercial Braille printers can be done away with, such as paper requirements, flattening of embossments leading to illegibility of characters, large amounts of noise being generated through printing (especially in solenoid embossment type printers). Therefore, substantial improvements can be carried out in the field of Braille printing through the technologies mentioned above.

## XII. Acknowledgements

The authors would like to thank Mr. Satyakam, Dept. of Electrical and Electronics, Manipal Institute of Technology, Manipal, for his support.

References

[1] Dandona L, Dandona R, John RK, "Estimation of blindness in India from 2000 through 2020: implications for the blindness control policy"

[2] American Foundation for the Blind "Cost of Braille Printers" (https://www.afb.org/ProdBrowseCatResults.asp?CatID=45)

[3] Ryles, Ruby. "The impact of braille reading skills on employment, income, education, and reading habits." Journal of Visual Impairment and Blindness 90 (1996): 219-226.

[4] US Patent 4183683 A – Line Printer for the raised-dot language of Braille Characters.

[5] US Patent 5527117 A - Braille printing solenoid housing.

[6] Patent WO 2011106762 A1 - Low-cost, portable, mechatronics-based-braille embossing apparatus and writing system for the blind.

[7] US Patent US 3598042 A – Braille printing System.

[8] US Patent US 4101688 A - Method for printing braille characters by lithography.

[9] I. Lita, D. Visan and A. Mazare, "Experimental Module for Assistive Technologies Applications", IEEE International Symposium for Design and Technology in Electronic Packaging, 2016.

[10] M. Kocidek, B. Wiqcek, G. De Mey, F. Steenkeste, C. Muntenau, "A Braille Printer On Reusable Thermoplastic Sheets", 21st Annual Conference and the 1999 Annual Fall Meetring of the Biomedical Engineering Society BMES/EMBS Conference, 1999.

[11] R Gupta, M. Balakrishnan, and P.V.M. Rao, "A Braille printer on reusable thermoplastic sheets" IEEE Potentials (Volume: 36, Issue: 1, Jan.-Feb. 2017).

[12] M. Nadeem, N Aziz et al., "A Comparative Analysis of Braille Generation Technologies", IEEE International Conference on Advanced Robotics and Mechatronics, 2016.

[13] R. Velasquez, E. Preza, H. Hernandez, "Making eBooks Accessible to Blind Braille Readers", IEEE International Workshop on Haptic Audio Visual Environments and their Applications, 2008.

[14] Stangl, Kim, Yeh, "3D printed tactile picture books for children with visual impairments: a design probe".

[15] P. Blenkhorn, "A system for converting braille into print," in IEEE Transactions on Rehabilitation Engineering, vol. 3, no. 2, pp. 215-221, Jun 1995.